\documentclass[sigconf,natbib=false]{acmart}
 \pdfoutput=1

\usepackage{glossaries}
\glsdisablehyper
\usepackage{siunitx}

\newcommand{\Watthour}{\si{\watt\hour}}

\newcommand{\minute}{\si{\minute}}
\newcommand{\second}{\si{\second}}
\newacronym[\glslongpluralkey={Non-Functional Properties}]{NFP}{NFP}{Non-Functional Property}
\newacronym{PCM}{PCM}{Palladio Component Model}
\newacronym{QoS}{QoS}{Quality of Service}
\newacronym{DVFS}{DVFS}{Dynamic Voltage and Frequency Scaling}
\newacronym{PSU}{PSU}{Power Supply Unit}
\newacronym{ACPI}{ACPI}{Advanced Configuration and Power Interface}
\newacronym{kWh}{kWh}{kilowatt hour}
\newacronym{VM}{VM}{Virtual Machine}
\newacronym{PDU}{PDU}{Power Distribution Unit}
\newacronym{SLA}{SLA}{Service Level Agreement}
\newacronym{IaaS}{IaaS}{Infrastructure as a Service}
\newacronym{PaaS}{PaaS}{Platform as a Service}
\newacronym{MAPE-K}{MAPE-K}{Monitor, Analyze, Plan, Execute, Knowledge}
\newacronym{S/T/A}{S/T/A}{Strategies, Tactics, Action}
\newacronym{OS}{OS}{Operating System}
\newacronym{HPC}{HPC}{High Performance Computing}
\newacronym{HDD}{HDD}{Hard Disk Drive}
\newacronym{SOA}{SOA}{Service-Oriented Architecture}
\newacronym{QuAL}{QuAL}{Quality Analysis Lab}
\newacronym{EE}{EE}{energy efficiency}
\newacronym{DES}{DES}{Discrete Event Simulation}
\newacronym{LQN}{LQN}{Layered Queueing Network}
\newacronym{IPMI}{IPMI}{Intelligent Platform Management Interface}
\newacronym{AIC}{AIC}{Akaike's Information Criterion}
\newacronym{RQ}{RQ}{Research Question}
\newacronym{UUID}{UUID}{Universally Unique Identifier}
\newacronym{PUE}{PUE}{Power Usage Effectiveness}
\newacronym{SPUE}{SPUE}{Server Power Usage Effectiveness}
\newacronym{PSM}{PSM}{Power State Machine}
\newacronym{FSM}{FSM}{Finite State Machine}
\newacronym{MARS}{MARS}{Multivariate Adaptive Regression Splines}
\newacronym{DLIM}{DLIM}{Descartes Load Intensity Model}
\newacronym{UPS}{UPS}{Uninterruptible Power Systems}
\newacronym{DC}{DC}{Direct Current}
\newacronym{AC}{AC}{Alternating Current}
\newacronym{ADL}{ADL}{Architecture Description Language}
\newacronym{SMM}{SMM}{Structured Metric Metamodel}
\newacronym{HRM}{HRM}{Hardware Resource Modeling}
\newacronym{DML}{DML}{Descartes Modeling Language}
\newacronym{PMX}{PMX}{Performance Model eXtractor}
\newacronym{SEFF}{SEFF}{Service Effect Specification}
\newacronym{SERT}{SERT}{Server Efficiency Rating Tool}
\newacronym{PCA}{PCA}{Power Consumption Analyzer}
\newacronym{EDP2}{EDP2}{Experiment Data Persistency \& Presentation}
\newacronym{PRM}{PRM}{Palladio Runtime Measurement Model}
\newacronym{QVTo}{QVTo}{Operational QVT}
\newacronym{ATL}{ATL}{ATL Transformation Language}
\newacronym{QVTr}{QVTr}{QVT Relations}
\newacronym{EMOF}{EMOF}{Essential Meta-Object Facility}
\newacronym{SD}{SD}{Story Diagram}
\newacronym{ERP}{ERP}{Enterprise Resource Planning}
\newacronym{MVC}{MVC}{Model-View-Controller}
\newacronym{EWMA}{EWMA}{exponentially moving weighted average}
\newacronym{SAS}{SAS}{Serial Attached SCSI}
\newacronym{TCO}{TCO}{Total Cost of Ownership}
\newacronym{MAE}{MAE}{Mean Absolute Error}
\newacronym{KDE}{KDE}{Kernel Density Estimation}
\newacronym{RDSEFF}{RDSEFF}{Resource-Demanding Service Effect Specification}
\newacronym{Wh}{Wh}{Watt hour}
\newacronym{REST}{REST}{Representational State Transfer}
\newacronym{IQR}{IQR}{interquartile range}
\newacronym{RT}{RT}{response time}
\newacronym{GQM}{GQM}{Goal Question Metric}
\newacronym{SPE}{SPE}{Software Performance Engineering}
\newacronym{EMF}{EMF}{Eclipse Modeling Framework}
\newacronym{UML}{UML}{Unified Modeling Language}
\newacronym{StoEx}{StoEx}{Stochastic Expressions}
\newacronym{FCFS}{FCFS}{first come, first served}
\newacronym{DSL}{DSL}{Domain Specific Language}
\newacronym{SECoMo}{SECoMo}{Software Eco-Cost Model}
\newacronym{eco-cost}{eco-cost}{ecological cost}
\newacronym{PET}{PET}{Performance counter Event Trigger}
\newacronym{CHAOS}{CHAOS}{Composable Highly Accurate OS-based power models}
\newacronym{FCO}{FCO}{Flexiant Cloud Orchestrator}
\usepackage{xifthen}
\usepackage{todonotes}
\usepackage[english]{babel}
\usepackage{multirow}
\usepackage{paralist}
\usepackage[utf8]{inputenc} 

\newif\ifproofread

\newcommand{\proofread}[1]{%
\ifproofread
\textcolor{blue}{#1}%
\else
\fi
}

\usepackage{booktabs} 

\setcopyright{none}

\acmDOI{tbd}

\acmISBN{tbd}

\acmConference[ICPE'18]{9th ACM / SPEC International Conference on Performance Engineering}{April 2018}{Berlin, Germany} 
\acmYear{2018}
\copyrightyear{2018}

\acmArticle{4}
\acmPrice{15.00}

\begin{document}
\title{Rapid Testing of IaaS Resource Management Algorithms via Cloud Middleware Simulation}
\subtitle{accepted paper, authors' preprint version for arXiv.org}

\author{Christian Stier}
\affiliation{%
  \institution{FZI Research Center for Information Technology}
  \city{Karlsruhe} 
  \state{Germany} 
}
\email{stier@fzi.de}

\author{J\"{o}rg Domaschka}
\affiliation{%
  \institution{Institute of Information Resource Management, Ulm University}
  \city{Ulm} 
  \state{Germany} 
}
\email{joerg.domaschka@uni-ulm.de}

\author{Anne Koziolek}
\affiliation{%
  \institution{Karlsruhe Institute of Technology}
  \city{Karlsruhe} 
  \state{Germany} 
}
\email{koziolek@kit.edu}

\author{Sebastian Krach}
\affiliation{%
  \institution{FZI Research Center for Information Technology}
  \city{Karlsruhe} 
  \state{Germany} 
}
\email{krach@fzi.de}

\author{Jakub Krzywda}
\affiliation{%
  \institution{Department of Computing Science\\ Ume\r{a} University}
  \city{Ume\r{a}} 
  \state{Sweden} 
}
\email{jakub@cs.umu.se}

\author{Ralf Reussner}
\affiliation{%
  \institution{Karlsruhe Institute of Technology}
  \city{Karlsruhe} 
  \state{Germany} 
}
\email{reussner@kit.edu}

\renewcommand{\shortauthors}{C. Stier et al.}

\begin{abstract}
\gls{IaaS} Cloud services allow users to deploy distributed applications in a virtualized environment without having to customize their applications to a
specific \gls{PaaS} stack.
It is common practice to host multiple \glspl{VM} on the same server to save resources.
Traditionally, \gls{IaaS} data center management required manual effort for optimization, e.g. by consolidating \gls{VM} placement based on changes in usage patterns.
Many resource management algorithms and frameworks have been developed to automate this process.
Resource management algorithms are typically tested via experimentation or using simulation.
The main drawback of both approaches is the high effort required to conduct the testing.  
Existing Cloud or IaaS simulators require the algorithm engineer to reimplement their algorithm against the simulator's API. Furthermore, the engineer
manually needs to define the workload model used for algorithm testing.
We propose an approach for the simulative analysis of \gls{IaaS} Cloud infrastructure that allows algorithm engineers and data center operators to evaluate optimization algorithms without investing additional
effort to reimplement them in a simulation environment. 
By leveraging runtime monitoring data, we automatically construct the simulation models used to test the algorithms.
Our validation shows that algorithm tests conducted using our \gls{IaaS} Cloud simulator match the measured behavior on actual hardware.
\end{abstract}

\maketitle

\section{Introduction}
\gls{IaaS} Cloud services allow users to deploy a distributed application in a virtualized environment without having to customize their application to a
specific \gls{PaaS} stack.
It is common practice to host multiple \glspl{VM} on the same server.
The shared hosting of \glspl{VM} reduces operational cost.   
Traditionally, \gls{IaaS} data center management required manual effort for optimization, e.g. by consolidating \gls{VM} placement based on changes in usage patterns.
Autonomic resource management addresses this problem by automating the allocation of virtual to physical resources. Resource management frameworks continuously optimize, e.g., the mapping of \glspl{VM} to servers. 
For this, they can leverage adaptation actions like \gls{VM} migration.

However, the design and selection of autonomic resource management algorithms for \gls{IaaS} data centers is a challenging task. 
In particular, the performance of the algorithms varies~\cite{manvi2014a}. Theoretical guarantees on the performance of resource management algorithms are only valid under impractical assumptions and thus cannot directly be used for the design and selection. The selection of resource management algorithms depends on the \gls{QoS} goals of an \gls{IaaS} data center operator, and tenant \glspl{SLA}. 
The selection of an algorithm thus requires an informed trade-off between conflicting goals~\cite{manvi2014a}. 

The experimental evaluation of algorithms, e.g. via benchmarking, requires large data center testbeds. This is both time consuming and costly.
\emph{Cloud simulators} like CloudSim \cite{Calheiros2011a} or GreenCloud \cite{Kliazovich2010a} offer reproducible conditions for algorithm testing. Once defined, it is possible to use the same workload scenarios to compare different resource management algorithms and configurations.

However, existing \gls{IaaS} Cloud simulators \cite{Calheiros2011a,Kliazovich2010a,Sakellari2013a} have specialized APIs, against which resource management algorithms need to be implemented. The re-implementation of algorithms for specific simulators is a challenging task. It requires expert knowledge of the simulator execution semantics, and their correspondence to the managed elements of the runtime management algorithms.
Changes made to the algorithm need to be implemented in the runtime and simulation variant of the algorithm. This induces significant effort.
Non-expert users, e.g. data center operators, have to rely on the availability of an algorithm implementation for their \gls{IaaS} simulator of choice.

Another difficulty of simulation-based evaluations is the acquisition of representative and accurate simulation models. 
The manual construction of simulation models, either in code, or graphical editors, requires significant effort. Furthermore, it requires detailed knowledge of the level of abstraction of the input model used by the simulator.

In this paper, we present an approach to integrate native resource management algorithm implementations into a data center simulation tool. The integration of native resource management algorithm implementations with the data center simulation has two main advantages. First, it removes the need to re-implement the algorithm against a simulation specific interface. This makes it easier to test resource management algorithms using simulation.
Second, an evaluation of the actual algorithm implementation increases confidence that it performs as intended. 

 We build upon our previous work, in which we suggested a generic approach to couple run-time models and simulation models~\cite{stier2016b}. 
The implementation of the integration approach leverages instances of the \emph{Adaptation Action} metamodel \cite{stier2016a} to define reusable and composable model-to-model transformation rules.

To address the challenge of simulation model acquisition, we present an automated simulation model extraction approach in order to reduce the effort for simulation model acquisition. 
Our approach reconstructs \gls{IaaS} workloads, including \gls{VM} submission and termination requests 
\proofread{over time} 
from historical measurements. Thereby, we enable the reconstruction of complex workloads. This enables the evaluation of runtime management algorithm performance under varying load.
We represent extracted \gls{VM} submissions using our timeline-based modeling language.
In a previous short paper \cite{krach2016b}, we introduced this language and an early evaluation of the language.
Algorithm engineers and data center operators can easily modify an extracted \proofread{timeline} model to evaluate alternative scenarios.

To summarize, the contributions of this paper are:
\begin{compactenum}
  \item An approach to integrate native resource management algorithm implementations into a data center simulation tool\label{contribution:1},
  \item A simulation model extraction approach that automatically reconstructs timeline-based, complex workload scenarios, \label{contribution:2}
	\item An evaluation of (\ref{contribution:1}), (\ref{contribution:2}), and the timeline-based modeling language \cite{krach2016b}.
\end{compactenum}


We evaluated our approach for a diverse set of \gls{IaaS} workloads. 
We used a set of scientific computing workloads to investigate the accuracy of extracted simulation models. 
The workloads were constructed based on expert knowledge on typical workloads submitted at the High Performance Computing Center at Ulm University.
We investigated the consistency of resource management decisions between simulation and a real world \gls{IaaS} testbed deployment at Ulm University.
We showed that our integrated approach accurately predicts data center utilization and power consumption metrics. 
Resource management decisions performed in the simulation are consistent with the behavior in the \gls{IaaS} testbed.
Finally, we illustrated the benefit of simulation-based testing for the selection of power management and autoscaling algorithms.
The simulation based evaluation of a power management algorithm showcased significant savings in energy consumption. Simultaneously, the algorithm did not compromise the deployment of \glspl{VM}.
The comparison of autoscalers enabled us to evaluate which algorithm better suited the scalability requirements for the investigated workload. 

This paper is organized as follows: Section \ref{sec:Foundations} describes the foundations of our work.
In Section \ref{sec:Example Use} we illustrate an example use case of our approach.
 Section \ref{sec:Coupling Middleware-Specific Runtime Management Implementations with Simulation} describes how native resource management algorithm implementations can be integrated into our data center simulation tool. 
Section \ref{sec:AutomatedSimulation} describes our simulation model extraction approach. The evaluation is presented in Section \ref{sec:eval}. Section \ref{sec:rw} compares our work to related work and Section \ref{sec:conclusion} concludes. 

\vspace{-4pt}
\section{Foundations}
\label{sec:Foundations}
\subsection{The CACTOS Project}
\label{sec:cactos}
The CACTOS project \cite{ostberg2014a} developed an approach for the autonomic management of \gls{IaaS} Cloud data centers.
As part of the CACTOS project, two toolkits were developed.
The \emph{CACTOS Runtime Toolkit} integrates monitoring and resource management via a variety of algorithms.
The \emph{CACTOS Prediction Toolkit} supports the systematic evaluation of alternative data center deployment scenarios.
This paper focuses on the prediction toolkit. 
\vspace{-1pt}
\subsubsection{CACTOS Runtime Toolkit}
The CACTOS Runtime Toolkit is designed to support different \gls{IaaS} Cloud platforms.
The toolkit integrates with these platforms to offer enhanced resource management capabilities.
As part of the project, OpenStack \cite{OpenStack} and \gls{FCO} \cite{FCO} support was developed.
CACTOS uses an optimization framework, CactoOpt \cite{ali2017predictive}, to derive adaptation action plans. 
The algorithms offered by CactoOpt \cite{ali2017predictive} aim for different \gls{QoS} trade-offs. In order to achieve these trade-offs, the algorithms formulate adaptation actions. They span the initial placement of \glspl{VM}, \gls{VM} migration, server level power management, and further resource management actions.
CactoOpt also offers autoscaling capabilities to horizontally scalable, multi-tier applications.
Autoscaling algorithms supported by CACTOS cover \emph{Hist}, \emph{ConPaaS}, \emph{Reg} and \emph{React} \cite{Ilyushkin2017a}.
The CACTOS Runtime Toolkit uses a runtime model, which was built specifically to automate the management of heterogeneous \gls{IaaS} data centers. The CactoOpt algorithms derive their plans from the runtime state that is represented in the runtime model.
A set of Cloud middleware components execute the adaptation actions in the data center environment.


\vspace{-3pt}
\subsubsection{CACTOS Prediction Toolkit}
The development, selection and pa\-ra\-me\-tri\-za\-tion of resource management algorithms are complex tasks.
The CACTOS project developed the CACTOS Prediction Toolkit to support what-if analyses for \gls{IaaS} Cloud data center environments. 
The CACTOS Prediction Toolkit builds upon the \gls{PCM} software performance model \cite{Becker2009a}, and the self-adaptive software systems simulator SimuLizar \cite{Becker2013b}. The toolkit supports the simulation of applications modeled at different levels of details, ranging from black-box \glspl{VM} to detailed application architecture models.
This paper contributes extensions to the toolkit that enable data center operators and algorithm engineers to systematically investigate the effectiveness and efficiency of resource management algorithms. Our work builds upon \cite{stier2016b,Svorobej2015a,Kistowski2017a}, which we discuss in the following.


\vspace{-3pt}
\subsection{Achieving Model Consistency between Runtime and Simulation Models}
\label{sec:Foundations:subsec:Consistency}
The model consistency approach described in \cite{stier2016b} leverages correspondence models and correspondence rules to synchronize runtime and simulation models.
The correspondence model holds the relationship between entities in the runtime and simulation model.
Correspondence rules can be subdivided in two categories. \emph{Mapping operations} synchronize the runtime model with changes in the simulation. Example changes are updates of measurements in simulation. The updated simulation measurements need to be propagated to the runtime model. This enables runtime management mechanisms to observe and react to changes in load.
\emph{Adaptation enactment rules} enact and synchronize the effect of adaptation decisions made by autonomic resource management mechanisms.
This paper applies \cite{stier2016b} to support the evaluation of resource management algorithms in the CACTOS Prediction Toolkit.

\subsection{Timeline-Based Experiment Scenarios}
\label{sec:Foundations:subsec:sec:Timeline Based Scenario Descriptions}
In order to assess the performance of \gls{IaaS} data center optimization algorithms, algorithm engineers and data center operators need workload models that are representative for the intended use cases of the algorithms.
Our \emph{Experiment Scenario} \cite{krach2016b} metamodel enables the specification of complex user interactions with data centers.
Instances of the metamodel represent interactions of users with a data center as a timeline of events.

\begin{figure}[htb]
	\centering
	\includegraphics[width=0.85\columnwidth]{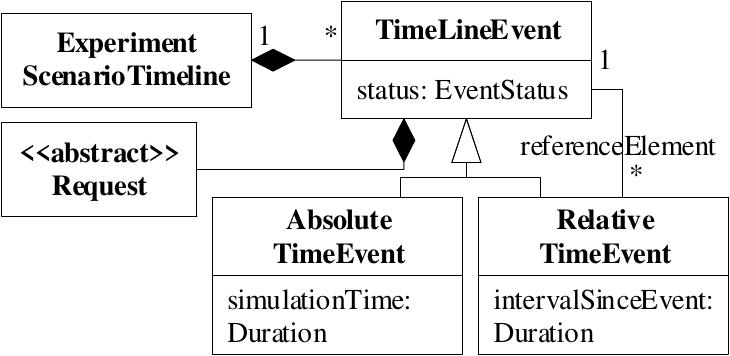}
	\caption{Excerpt from the Experiment Scenario metamodel.}
	\label{fig:ExperimentScenarioMetamodel}
\end{figure}

Figure \ref{fig:ExperimentScenarioMetamodel} depicts the central classes from the metamodel. The \emph{ExperimentScenarioTimeline} consists of \emph{TimeLineEvents}. 
Each event maps a \emph{Request} model element to the timeline. 
There are two types of events. \emph{AbsoluteTimeEvent} specifies an absolute point in time at which a request should be triggered. \emph{RelativeTimeEvent} defines the request time relative to another event. An example request type is \emph{StartApplicationRequest}. 
\emph{StartApplicationRequest} models a user request to start a new individual \gls{VM} or distributed application. The request references an application template that is used to assemble the application.
Reconfigurations of the data center caused by manual intervention of operators can be expressed with appropriate requests. 
For an overview of all supported types we refer to \cite{krach2016b}.

\proofread{The events on the timeline are not simulated directly. Rather, they are transformed into \gls{PCM}~\cite{Becker2009a} model elements, and events in the SimuLizar \cite{Becker2013b} simulator. VM workloads are represented as \emph{Assemblies}, \emph{Usage Scenarios} and \emph{Usage Evolutions}~\cite{groenda2015c}.}

In simulation, a dedicated event scheduler processes the \emph{AbsoluteTimeEvents} ordered by ascending simulation time. The scheduler tracks the execution status of events in their \emph{EventStatus}. Adaptation enactment rules trigger the execution of the requests. 
\proofread{Here, we reuse existing rules for the enactment of adaptation decisions, as Section \ref{sec:Coupling Middleware-Specific Runtime Management Implementations with Simulation:subsec:InformationGap} explained. 
The event handler triggers the execution of the adaptation enactment rules linked with the scheduled \emph{Request}.}
\proofread{The scheduler supports the consideration of execution times for the user requests. For \emph{StartApplicationRequest}, the execution time depends upon the execution time of the \gls{VM} placement algorithm, and the time required for the placement itself. 
In order to model these overheads, we employ reusable adaptation performance models, as described in our previous work \cite{stier2016a}.}
\proofread{Following the execution of the request during simulation, the workload descriptions associated with the respective VMs are executed.} 

\proofread{
User actions, such as VM boot-ups, take time to execute. 
\emph{RelativeTimeEvents} execute dependent on the completion time of their \emph{referenceEvent}.
This enables the evaluation of orchestrated deployments, where a user deploys \glspl{VM} in a specific order.}

\proofread{
Simulated VM workload is issued to the system via the VM's Usage Scenario, as described in ~\cite{groenda2015c} for black-box applications. We have extended SimuLizar to support the addition removal of VMs, by adding support for a varying number of active Usage Scenarios. During simulation, new users (the workload driver) are generated automatically, after the respective Usage Scenario is added to the active Usage Model by the request processing transformation. Similarly, a VM stop request is represented by a removal of the representing PCM elements, particularly the Usage Scenario, which in turn prevents new users from being issued. Users which have already started execution of the scenario run to completion before being removed.	
}

\begin{figure}[htb]
	\centering
	\includegraphics[width=\columnwidth]{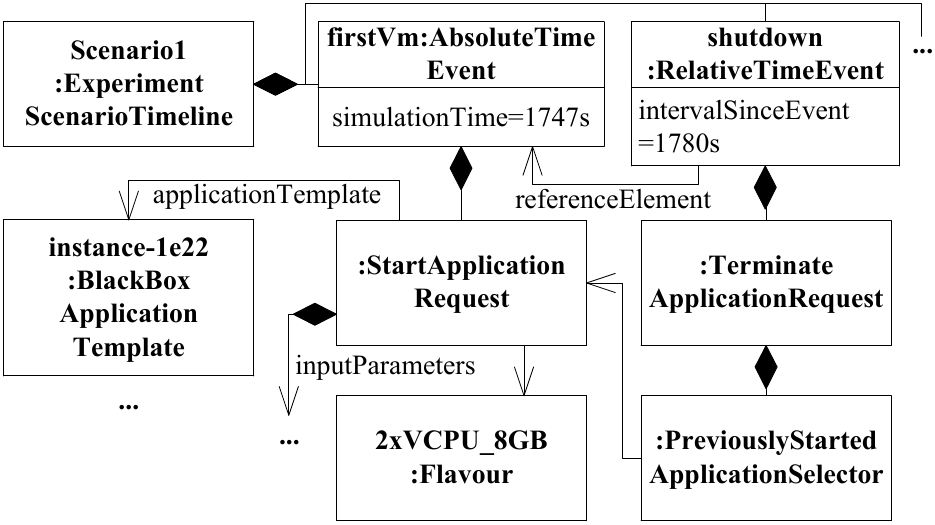}
	\caption{Excerpt from example Experiment Scenario model.}
	\label{fig:ExperimentScenarioExample}
\end{figure}

Figure \ref{fig:ExperimentScenarioExample} depicts an example excerpt from an Experiment Scenario model based on one of our evaluation scenarios. It contains a start-up request for the \gls{VM} \emph{instance-1e22}. The scenario prescribes that \emph{instance-1e22} should be started at simulation time $1747\second$, and terminated $1780\second$ later.
The startup request references the \gls{VM} template, used \gls{VM} flavor and input parameters.
The terminate request references the prior \emph{StartApplicationRequest}, as the \gls{VM} to be terminated is not yet running in the initial simulation model.

\subsection{Extracting Data Center Simulation Models}
\cite{Svorobej2015a} present an initial approach that leverages a runtime model snapshot from the CACTOS Runtime Toolkit as foundation for simulations. 
\proofread{The runtime model gives an accurate representation of the current data center state.} 
The runtime model lacks information on historically executed \glspl{VM} and their workloads. 
The runtime model thus can only be used to evaluate how runtime management algorithms would perform under stable load conditions. 
\cite{Kistowski2017a} sketches an algorithm for the reconstruction of black-box resource demand functions from a series of load measurements. 

This paper contributes a novel model extraction approach that supports the reconstruction of timeline-based workload models from historical measurements. 
It leverages \cite{Svorobej2015a} to gather basic infrastructure information, i.e., on the available servers.
Our model extraction approach applies the algorithm from \cite{Kistowski2017a} to reconstruct workload models for individual \glspl{VM}.

\section{Example Use}
\label{sec:Example Use}
A data center operator might use our approach as follows. As a starting point, she might be interested in evaluating how the introduction of automated resource management would affect the performance and efficiency of a manually managed data center.
For this, the data center operator can install the monitoring tools provided by the CACTOS Runtime Toolkit to gather monitoring data.
Next, the operator applies our simulation model reconstruction methods to the data.
The operator then can simulate how the application of an existing runtime resource management algorithm implementation would have affected the performance and efficiency of the data center for this scenario under investigation.

The use of existing algorithm implementations and automated model construction significantly reduces the evaluation effort for the data center operator. It rules out inconsistencies between simulation and runtime implementation variants. 
It thus increases confidence in the simulation results.
Once she has invested the initial effort for the monitoring setup, the operator can continuously reevaluate and compare different algorithms. This enables the operator to adapt the algorithm choice and configuration to changes in the data center setup, workload and performance requirements.

\section{Integrating Middleware-Specific Runtime Management Algorithms with Simulation}
\label{sec:Coupling Middleware-Specific Runtime Management Implementations with Simulation}
Runtime management frameworks  use \emph{runtime models} to manage resources. Runtime management algorithms leverage information from the runtime models to plan adaptation decisions.
Existing Cloud and simulation frameworks lack support for simulating these algorithms in their middleware-specific implementation. 
This section presents our approach for the integration of middleware-specific runtime management algorithms with an \gls{IaaS} Cloud simulator.
It enables algorithm engineers and data center operators to test and evaluate algorithms, while requiring minimal knowledge of the simulation API.

	\subsection{Information Gap between Runtime Models and Simulation Models}
\label{sec:Coupling Middleware-Specific Runtime Management Implementations with Simulation:subsec:InformationGap}
Software system simulators like SimuLizar \cite{Becker2013b} or CloudSim \cite{Calheiros2011a} naturally abstract from information that is not needed to predict the metrics which are relevant to the use cases of the simulator.
The abstraction covers characteristics of the hardware and software stack.
This simplifies the simulation, as well as the construction of input models for simulator users.

The level of abstraction chosen when modeling individual entities depends on the pragmatism of the simulation. 
Many software performance simulators do not model memory \cite{Becker2013b,Becker2009a,Kliazovich2010a}, as (i) memory accesses are difficult to predict, and (ii) their effect on \gls{QoS} is considered negligible for CPU-bound applications.

Runtime models are designed to support autonomic resource management.
This contrasts the pragmatism of design time performance models like \gls{PCM}~\cite{Becker2009a}. Design time \proofread{performance} models focus on modeling system characteristics that impact performance.
Runtime models capture all characteristics which are relevant to the management of a system. In \gls{IaaS} data centers, this can include user management information and detailed \gls{VM} instantiation parameters.
Unlike design time performance models, runtime models may not capture information on user and application behavior on a level that is detailed enough for performance simulations.

A naive approach is to transform the runtime models to performance models of the simulator. 
This approach, however, requires that all resource management algorithms are reimplemented against the model of the simulator. Runtime management algorithms, which consider properties that are not reflected in the simulation model, can not be simulated.
We designed an approach for achieving model consistency between runtime models and simulation models. Our approach supports simulation-based analysis and testing of optimization algorithms without the need to modify or re-implement the algorithms.

\subsection{Achieving Model Consistency}
\label{sec:Coupling Middleware-Specific Runtime Management Implementations with Simulation:subsec:Coupling}

\proofread{An essential advantage of the CACTOS Prediction Toolkit compared to other Cloud simulators \cite{Calheiros2011a,Kliazovich2010a} is the ability to evaluate resource management algorithms \emph{without modification}.} 
We achieve model consistency by implementing the approach discussed in Section \ref{sec:Foundations:subsec:Consistency} to achieve model consistency between the CACTOS runtime model and the \gls{PCM} simulation model. A specialized metamodel maintains the correspondence between runtime and simulation model.
In total, the metamodel distinguishes 40 correspondence types.

\begin{figure}[htb]
	\centering
	\includegraphics[width=\linewidth]{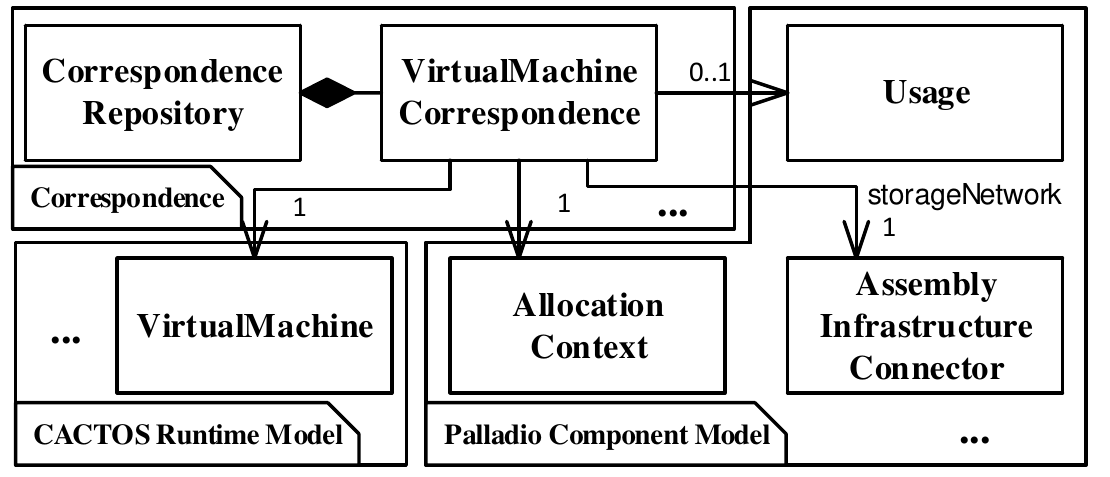}
	\caption{Excerpt from the correspondence metamodel between runtime and simulation model.}
	\label{fig:correspondenceExample}
\end{figure}

Figure \ref{fig:correspondenceExample} provides an example correspondence from this metamodel. Each \emph{VirtualMachine} in the CACTOS runtime model corresponds to multiple entities in the simulation model, the \gls{PCM}.
 The \emph{AllocationContext} represents the deployment of a set of software components to a server. It can be used to represent \gls{VM} allocations in \gls{PCM}. However, the correspondence with allocated components fails to cover all aspects that are needed to enable meaningful simulations. The CACTOS runtime model contains detailed storage information of \glspl{VM}. This includes, e.g., the location of their network attached devices. In order to reason the effects of remote storage accesses, the storage characteristics need to be mapped to \gls{PCM}.
Unlike the runtime model, the \gls{PCM} explicitly models user interactions with the \gls{VM} components. The correspondence covers this with the reference to the \emph{Usage} model.

A model-to-model transformation from the CACTOS runtime model
to \gls{PCM} establishes the initial correspondence between runtime and simulation model. The mapping operations update the correspondence model during simulation.
We implemented the operations as modular model-to-model transformations for the model-based simulated runtime management of SimuLizar \cite{Becker2013b,stier2016a}. The implementation uses the Adaptation Action metamodel \cite{stier2016a} to specify the mapping operations in a reusable manner.

\proofread{The correspondence model needs to maintain all information that can not be derived from the runtime model or simulation model using rules. The mapping of remote storage is an example of this. Without the correspondence model, it would be impossible to infer which remote storage connector in the \gls{PCM} belongs to a \gls{VM}.}

\subsection{Implementation of Runtime Management Integration}
We integrated all runtime management algorithm types supported by the Cloud middleware of the CACTOS Runtime Toolkit with the simulation environment of the CACTOS Prediction Toolkit. This enabled us to support all algorithms of the same type. No additional effort was required to integrate specific algorithms with the simulator.

A \emph{Placement Connector} bridges the gap between the simulated runtime management, and the placement algorithms of CactoOpt.
When a new \gls{VM} startup request is issued, the simulated runtime management calls the native optimization implementation via the placement connector.
The model consistency mechanism \proofread{we} discussed in Section \ref{sec:Coupling Middleware-Specific Runtime Management Implementations with Simulation:subsec:Coupling}
\proofread{implicitly} realizes the simulated runtime management. \proofread{Thus, it was unnecessary to invasively extend the simulator.}
		
We integrated data center optimization algorithms via an \emph{Optimization Connector}. This connector couples the resource management optimization algorithms and heuristics of CactoOpt with the simulation.
The Optimization Connector functions similar to the Placement Connector. 
Instead of a single supported decision, i.e., placement, the consistency mechanism translates a wide range of server and \gls{VM} reconfiguration decisions to simulation. It uses the adaptation enactment rules for this purpose.


The \emph{Autoscaling Connector} communicates with the native autoscaling middleware of CACTOS. The middleware manages autoscaling on a per-application basis. It uses the runtime model passed by the simulation to decide on horizontal scaling.
\proofread{The current simulated load is particularly relevant to evaluate the performance of autoscalers.} 
Instead of measurements from a real world data center, the simulation runtime management exposes simulated measurements to the autoscalers. Mapping operations of the rule-based synchronization engine update the measurement representations in the runtime model, as Section \ref{sec:Coupling Middleware-Specific Runtime Management Implementations with Simulation:subsec:Coupling} described.

\proofread{Reconfigurations of the data center caused by manual intervention of system administrators can be expressed with appropriate requests. 
\emph{ReconfigureOptimisationAlgorithmRequest} supports a switch to a different optimisation algorithm during the simulated scenario. \emph{ChangeOptimisationIntervalRequest} changes the frequency with which the autonomic resource management algorithms execute.
The two actions enable data center operators to evaluate if a data center would gracefully switch between two sets of runtime management configurations. For an overview of all supported types we refer to \cite{krach2016b}.}

\section{Automated Simulation Model Construction}
\label{sec:AutomatedSimulation}
	A major challenge in the application of simulations is the acquisition of simulation models. It is more attractive to reason on the performance of resource management algorithm using simulation if the models can be obtained with little effort.
We implemented an approach that enables the evaluation of resource management algorithm performance using automatically constructed performance models.
Our approach uses the historical information collected by the CACTOS Runtime Toolkit to construct performance models.

\subsection{Black-Box Performance Models}
\label{sec:AutomatedSimulation:subsec:ModelReconstruction:subsubsec:BlackBox}
Black-box \gls{VM} workload models describe \gls{VM} workloads in terms of their resource usage over time.
The main benefit of black-box VM models is that they do not require insight into the applications deployed in a VM. As they only depend upon metrics and topology information available to the data center operator, they can be constructed for any VM based on its past observed behavior.
We construct the black-box models from past load measurements, which were recorded in the historical database of the CACTOS Runtime Toolkit.
We normalize the load levels of \glspl{VM} using the processing speed of their original host to account for \gls{VM} migrations.

\subsection{Timeline-Based Experiment Scenarios}
\label{sec:AutomatedSimulation:subsec:ModelReconstruction:subsubsec:ExperimentScenarios}
Data center operators and algorithm engineers can use instances of the Experiment Scenario model to evaluate complex user interaction with the simulated data centers. Section \ref{sec:Foundations:subsec:sec:Timeline Based Scenario Descriptions} introduced the Experiment Scenario metamodel, and provided an overview of supported user interactions, e.g., \gls{VM} submissions. 
The manual modeling of Experiment Scenarios requires expert knowledge of potential \gls{VM} submission patterns, typical \gls{VM} configurations, and workloads.
We realized an automated approach for the reconstruction of Experiment Scenarios from historical measurements. It enables data center operators and algorithm engineers to evaluate resource management algorithms based on past workloads and user interactions.

Our approach uses recorded \gls{VM} submission and reconfiguration events to reconstruct an Experiment Scenario timeline.
We link each \gls{VM} submission to a black-box performance model, which is automatically constructed.
In order to construct an Experiment Scenario model, the user specifies a period of time, and a subset of servers she is interested in.
We translate this to a set of queries on a historical measurement database. The data from these queries is funneled into the reconstruction of user interactions.
\proofread{\gls{VM} submissions and shutdowns that are not submitted by data center tenants, but by autoscalers, can be filtered from the results.}
We enrich the Experiment Scenario with \gls{VM} instantiation parameters. This increases the accuracy of placement decisions in simulation. Placement algorithms \cite{ali2017predictive} consider these parameters to determine if \glspl{VM} can be deployed on the same server without causing resource contention.

\subsection{Power Models}
\label{sec:AutomatedSimulation:subsec:ModelReconstruction:subsubsec:Power Models}
Power consumption decisively determines the operational cost of data centers.
System-level power models \cite{Rivoire2008a} enable power consumption predictions of individual servers based on metrics like CPU utilization.		
Power models need to be trained for specific servers in order to make accurate power consumption predictions. 
We realized an approach that uses the historical data collected in the historical measurement database as the source of training data for statistical power model learning.
For a given time frame, we query the collected measurements. We use this data as input to power model training. 
A non-linear regression technique 
\proofread{
\cite{Rousseeuw2016a}} trains a given power model type on the selected training data set. 
\proofread{An example power model type is 
\begin{center}
\begin{math}
P(u)=c_0\cdot u+c_1 \cdot u^2+c_2 \cdot u^3+c_3
\end{math},
\end{center}
where u is the aggregate CPU utilization at a point in time.
Here, the regression technique trains the constants $c_0$ through $c_3$.}

\proofread{
We clean the input data prior to training.
For each discrete utilization metric value, we aggregate all collected power consumption measurements for that value. This reduces the effect of noise and consumption fluctuations on the trained regression model.}
\subsection{Limitations}
The reconstruction of black-box \gls{VM} and timeline-based scenario models is well suited to evaluate the performance of resource management algorithms for workloads observed in the past. It is of limited use to explore scenarios that involve user-facing applications with varying workloads. For this, other means of performance model acquisition are more suited.
The extraction of server power models from historical measurements requires that the server has run workloads which cover the utilization ranges investigated in simulation.

\section{Evaluation}\label{sec:eval}
In order to evaluate our approach, we compared simulation results and measurements for a set of experiments conducted in a data center testbed.

\subsection{Scientific Computing}
We evaluated the applicability of our approach using a case study from the scientific computing domain.

\subsubsection{Scenario Description}
We conducted the case study in a commodity hardware testbed.
The testbed was operated using the OpenStack Cloud middleware in combination with the CACTOS Runtime Toolkit. The servers ran KVM hypervisors. The CACTOS Runtime Toolkit contributed autonomic resource management.
In each of the evaluated configurations, \gls{VM} placement and migration algorithms were in use.
We used IPMI to collect power consumption measurements of the servers over time. 

In order to evaluate the accuracy of our integration and model extraction methods, we proceeded as follows. First, we ran an experiment in an \gls{IaaS} data center testbed.
Second, we obtained a simulation model by applying our model extraction method. We applied our Experiment Scenario extraction to obtain models of the user interactions, and \gls{VM} black-box workload models from the experiment run.
Next, we conducted a simulation using the resulting input models.
We used the same algorithm implementations and configurations for the simulated run as in the testbed experiment run.
Finally, we compared measured and predicted results.

The following outlines the scenarios for which we conducted the experimental evaluation. Every scenario encompassed a set of scientific computing workloads. Specifically, we executed a set of Molpro \cite{Molpro} workloads. Molpro is a framework for quantum chemistry calculations. 
Molpro follows run-to-completion semantics, as is common for scientific computing applications.
A compute job submission system translated each scientific compute job submission request to a \gls{VM} submission request on the \gls{IaaS} testbed.
Per scenario, a load driver submitted the jobs over time based on a predefined submission schedule.  
We constructed the submission schedule to resemble a typical daily cycle of user job submissions in the High Performance Computing Center at Ulm University.
The job submissions consist of a mix of long and short running jobs. The short jobs reach execution times of up to two hours. Long running jobs may span eight to ten hours. 

\paragraph{Scenario 1} The first scenario covered 26 \gls{VM} submissions to a testbed setup which consisted of eight servers. 
Six of these eight servers had a power meter, from which we could collect measurements.
The experiment lasted just short of one and a half hours.
We used consolidation algorithms for both \gls{VM} placement and migration. They consolidated the \glspl{VM} based on their RAM requirements.

\paragraph{Scenario 2} The second scenario consisted of 15 compute job submissions. It covered a run time of approximately eight and a half hours. 
We allocated the same eight servers as Scenario 1. 
We configured the Runtime Toolkit to use load balancing algorithms for both \gls{VM} migrations and placement. The algorithms aim to evenly distribute the \glspl{VM} on all servers based on their RAM requirements.

\paragraph{Scenario 3} Scenario 3 encompassed 19 compute job submissions. It covered the same basic experiment as Scenario 1, but with an extended run time of eight hours and 46 minutes. The 26 \gls{VM} submissions from Scenario 1 were reduced to 19. The \glspl{VM} were hosted on the same set of eight servers. We used RAM based consolidation algorithms for \gls{VM} placement and migrations. 

\paragraph{Scenario 4} The fourth scenario consisted of 37 Molpro job submissions. It covered a run time of roughly 26 hours. 
It used six servers from the \gls{IaaS} testbed. We could collect power measurements from four of these six servers.
Scenario 4 used the same migration and placement algorithms as Scenario 1 and 2. 

\subsubsection{Results}

\proofread{The following compares the measured and simulated behavior of the scenarios.
First, we investigate if our model extraction and algorithm integration approach results in an accurate prediction of placement decisions for Scenario 1. Second, we compare the overall accuracy for each of the scenarios by comparing total measured and predicted energy consumption.
}




For  Scenario 1, the algorithms placed the \glspl{VM} on the same servers as in the measured experiment. 
In order to quantify the prediction accuracy over the duration of the experiment, we compared the predicted and measured accumulated energy consumption of all servers with power meters
 using the error formula
\begin{math}
|\frac{E_\text{Meas}-E_\text{Sim}}{E_\text{Meas}}|
\end{math}, where $E$ is the aggregate energy consumption.

\begin{table}[htbp]	
	\caption{Total measured and predicted energy consumption for the four evaluated scenarios, with prediction error. Duration in minutes. Energy consumption in $\Watthour$, error in $\%$.}
	\label{tab:comparisonPowerPetClinicMs}
	\centering
	\begin{tabular}{l r r r r}
		\toprule
		Scenario & Duration & Measured & Predicted & Error \\ \midrule
		1 & $75\,\minute$ & $1\,783\,\Watthour$ & $1\,661\,\Watthour$ & $6.85\%$ \\
		2 & $514\,\minute$ & $5\,443\,\Watthour$ & $5\,464\,\Watthour$ & $0.39\%$ \\
		3 & $526\,\minute$ & $5\,238\,\Watthour$ & $5\,609\,\Watthour$ & $7.08\%$ \\
		4 & $1561\,\minute$ & $13\,558\,\Watthour$ & $12\,826\,\Watthour$ & $5.40\%$ \\
		\bottomrule
	\end{tabular}
\end{table}

\proofread{The energy consumption prediction reached an error of less than 7\% compared to the measured consumption.
}
In Scenario 1 we employed a linear power model to predict the energy consumption of the servers. 
The linear model was trained using historical measurements from each server.
We used more complex power models for the other scenarios, e.g., with exponential components.
Table \ref{tab:comparisonPowerPetClinicMs} lists the measured and predicted total energy consumption over each run. In Scenario 2, the energy consumption prediction reached a prediction error of $0.39\%$. Scenario 3 had the highest prediction error at $7.08\%$.

We could trace back the source of the prediction error for Scenario 3 to a lack of historical measurements from one of the \glspl{VM}. While the \gls{VM} was running in the experiment, its measurements were not recorded due to the failure of \gls{VM} internal monitoring.
Thus, our tooling was unable to reconstruct a behavior model of the \gls{VM}. \proofread{Consequently, the experiment scenario model used in the simulation did not contain the placement of the \gls{VM}.
This impacted placement decisions in the simulation.
The missing \gls{VM} left capacity on one of the servers without power meters.} Down the line, this led to the placement of a highly active \gls{VM} on one of the servers with a power meter. This increased the predicted energy consumption. \proofread{In the measurement run, the consolidation algorithm placed this \gls{VM} to a server without a power meter due to RAM constraints.}

\subsection{Power Management}
In order to validate that our algorithm integration approach also supports the analysis of power management algorithms, we applied an existing algorithm to the Scenario 3 workload. We configured \gls{VM} migration and placement to consolidation algorithms. This enabled the power management algorithm to turn off free servers. Our simulations were able to show significant power savings, without negatively affecting the deployment of new \glspl{VM}.

\subsection{Autoscaling}\label{sec:eval:subsec:Autoscaling}
The selection of the right autoscaling algorithm for an application is a challenging task. 
This section explores how we can employ our simulation-based method to compare different autoscaling policies for an enterprise web application. 

\subsubsection{Compared Autoscalers}
\proofread{In order to offer improved scalability, w}We evaluated which of two autoscalers performed the best for the evaluated enterprise application. We compared the two autoscalers \emph{React} and \emph{Reg}.

React \cite{Chieu2009a} is a rule-based autoscaler. 
Its algorithm increases the number of active instances of a scalable application tier if the measured user workload surpasses a specified threshold capacity. If at least two instances are under-utilized, React shuts down and decomissions one instance.

Reg \cite{Iqbal2011a} is an autoscaler, which scales the number of active instances based on a regression model.
If the measured load falls below a specified threshold, the autoscaler reduces the number of active instances. It uses a regression model to determine the number of active instances, which should remain active.
For user workload levels higher than a threshold capacity, Reg initiates the startup of additional instances.

\subsubsection{Case Study System}
DataPlay\footnote{\url{https://github.com/cactos/DataPlay}, last retrieved 24.10.2017.} is a horizontally scalable multi-tier enterprise web application. It is a gamified social platform for data exploration. \proofread{, which is available as open source.} 
DataPlay follows a three-tier architecture style, where the business tier can be horizontally scaled.



We obtained the input model for our simulation by enriching a runtime model snapshot of the data center. The snapshot contained a description of its server infrastructure, and an application model \proofread{template} of DataPlay. \proofread{The CACTOS Runtime Toolkit instantiates this template when a user deploys DataPlay in the runtime environment. The toolkit attaches the white box application model description to the instances of DataPlay \glspl{VM}.}
We instantiated the application at the beginning of the simulated experiment using our Experiment Scenario model.

We used a synthetic workload that covered a wide range of workload intensities, and workload variations. The workload covered a time frame of over 
eleven and a half hours of simulation time. Over this period, a seasonal pattern repeated sixteen times. The workload reached approximately 100 requests per second at its peak. It contained short periods with request rates just above, or at zero. The seasonal pattern was folded with uniform noise in the interval $[-3,2]$ requests per second.

\begin{figure}
\includegraphics[width=\columnwidth]{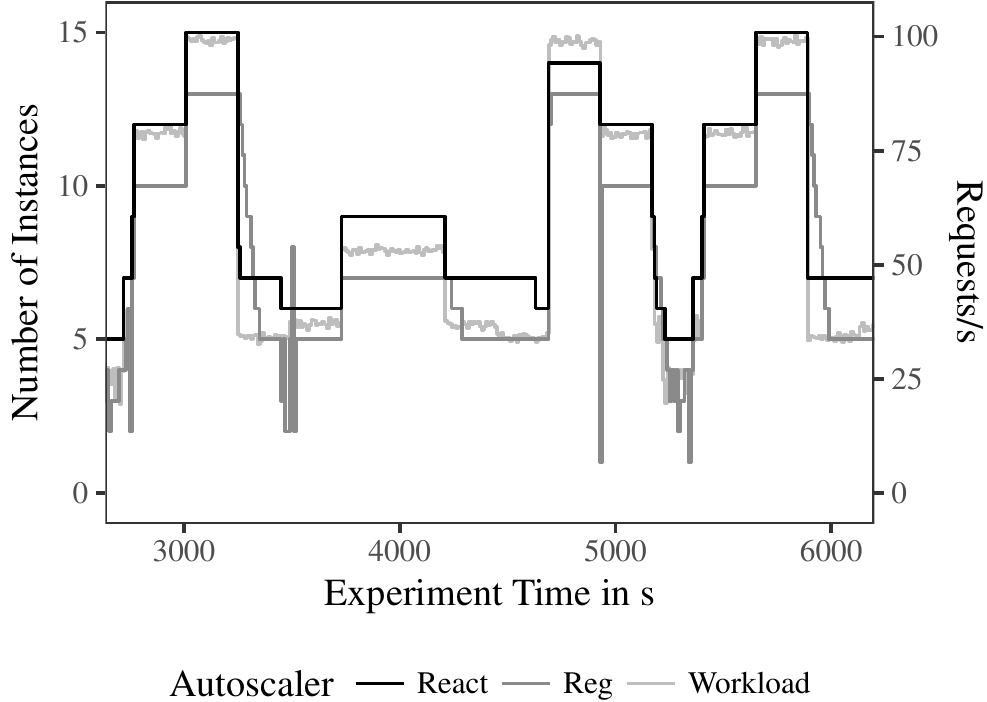}
\caption{\label{fig:AutoscalingVMs}Experimental dynamics of the two simulated autoscaling policies. Excerpt from the full experiment, which spanned a 
\proofread{simulated} 
time frame of over 11 hours. The black line shows the workload intensity as requests per s. The grey lines represent the number of active instances over time, which the autoscalers allocate.}
\end{figure}

\subsubsection{Results}\label{sec:eval:subsec:Autoscaling:subsubsec:Results} We look at an excerpt of the whole experiment results to discuss our findings.
Figure \ref{fig:AutoscalingVMs} illustrates the experimental dynamics of the autoscalers and the workload. The \emph{workload} line shows the rate at which requests arrived at the DataPlay application.
The grey lines represent the total number of \glspl{VM} over time, which the autoscaling algorithms suggested to keep allocated. 
The behavior of both algorithms differed significantly.
Reg frequently triggered scale-out and scale-in decisions. Particularly, Reg over-eagerly performed scale-ins once the workload started to decrease. This led to staggered scaling, e.g., at around 3500 experiment time.
React over-provisioned \glspl{VM}. Compared to Reg, it allocated more or an equal number of \glspl{VM} most of the time.
React recommended to operate $9.12$ \glspl{VM} on average, while Reg only proposed $7.47$.
Over the course of the experiment, Reg issued 2544 scale-in or scale-out actions, while React only adapted 647 times. The higher frequency of Reg led to a larger overhead for (de-)commissioning \glspl{VM}.

The poor performance of Reg is in line with the experimental comparison of autoscalers by\cite{Ilyushkin2017a}. In their experiments, Reg also under-provisioned \glspl{VM} and quickly varied the number of active \glspl{VM}.
The authors only could improve the performance of Reg, once they implemented a set of improvements to the original algorithm and its implementations.
React manages to match demand in most periods, or overprovisions.

In conclusion, we determined that React provides better operational stability at the cost of light overprovisioning. Thus, we consider React to be better suited as a autoscaler for DataPlay for the investigated workload patterns.
We did not record performance metrics, such as response time, and average, minimum and maximum CPU utilization as part of our comparison. In future work, we plan to compare the autoscaling policies based upon these further metrics, and the metrics outlined in \cite{Ilyushkin2017a}.

\section{Related Work}\label{sec:rw}
The simulation-based evaluation of Cloud resource management has been a topic of great interest in recent years.
\cite{Sakellari2013a} provide an overview of Cloud simulators.
Two popular \gls{IaaS} Cloud simulators discussed by the survey are CloudSim~\cite{Calheiros2011a} and GreenCloud~\cite{Kliazovich2010a}. Both support the evaluation of resource management algorithms. However, they require a reimplementation of the algorithms for the simulator specific APIs.

\cite{Vondra2017a} present a Cloud simulator that has been built for the simulation-based evaluation of autoscaling algorithms. Like CloudSim and GreenCloud, their simulator requires a reimplementation of each algorithm for the simulator interface.

CDOSim \cite{Fittkau2012a} extends CloudSim~\cite{Calheiros2011a} to support the evaluation of enterprise Cloud application migration scenarios. CDOSim offers an approach to extract white-box application models using static code analysis, and dynamic instruction counting. This requires source-code level access to, and extensive profiling of the evaluated Cloud application. In the context of our work, it could be applied to extract white-box application models.

\cite{Calheiros2013a} propose a profiling based approach to construct coarse grey-box workload models of applications. Their approach requires dedicated profiling infrastructure. It profiles the Cloud application with varying workload intensities and compute resources. This complements our black-box model extraction approach. Unlike our approach it can, however, not be applied to model arbitrary \gls{VM} workloads. 

\cite{Ilyushkin2017a} experimentally evaluate a set of seven state of the art autoscaling algorithms. The authors evaluate the algorithms for scientific computing workloads. The comparison required a complex \gls{IaaS} testbed setup and extensive experiments. Our work aims to reduce the effort for testing using simulations. Indeed, we were able to evaluate two of the algorithms from \cite{Ilyushkin2017a}, of which we had the implementations.

\section{Conclusion}\label{sec:conclusion}
This paper presents an approach for rapid testing of resource management algorithms for \gls{IaaS} Cloud data centers.
Our approach enables algorithm engineers and data center operators to evaluate \gls{IaaS} Cloud resource management algorithms using simulations. Our simulation-based approach supports the simulation of algorithms which are natively implemented for Cloud middleware.
The CACTOS Prediction Toolkit implements our approach for the CACTOS Runtime Toolkit, and its supported adaptation actions.
These actions include the initial placement of \glspl{VM}, \gls{VM} migration, power management and autoscaling.
We show that our integration approach enables reasoning on the performance of diverse types of resource management algorithms.

\proofread{The CACTOS Prediction Toolkit facilitates the combination of different simulation models. It simulates \gls{VM} application workloads modeled at different levels of abstraction.
We outline how these models can be constructed from runtime models and historical measurements.}

We evaluated our approach for a diverse set of real-world workloads. We evaluated a set of resource management algorithms for scientific computing application workloads.
The results from simulation have a high accuracy for energy consumption and utilization measurements. \gls{VM} placement and migration decisions are consistent between the measured and simulated experiments.
Our simulation enabled us to explore the effect of active power management algorithms on total consumption, and the reliability of \gls{VM} placements. Thereby, the simulation-based evaluation helps avoid scenarios where power management interferes with the ability of a data center to serve all \gls{VM} submission requests.
We applied our approach to compare two autoscalers for an enterprise web application. Our observations on the autoscaler dynamics from simulation are consistent with published experimental evaluations \cite{Ilyushkin2017a}.

Our approach enables algorithm engineers and data center operators to rapidly evaluate the performance of \gls{IaaS} resource management algorithms.
It requires no additional effort or in-depth knowledge of simulation models and APIs.
The users of our approach can simply evaluate their existing resource management algorithm implementations.
We automate the construction of simulation models. For this, we leverage existing runtime models and historical measurements.

We plan to expand our approach in two directions. First, we aim to automate the construction of detailed application models of scientific computing applications. This will reduce the effort for the construction of accurate simulation models, which consider the phases of scientific computing applications.
Second, we plan to conduct case studies which investigate the prediction accuracy of application workloads with large heterogeneity between workloads, and used servers. We intend to expand the quantitative comparison of simulation and measurements of autoscaling policies to the metrics listed in Section \ref{sec:eval:subsec:Autoscaling:subsubsec:Results}.

\appendix
\begin{acks}
This work is funded by the European Union's Seventh Framework Programme under grant agreement 610711 (CACTOS), the Swedish Research Council (VR) project Cloud Control and the Swedish Government's strategic research project eSSENCE.
\end{acks}



\end{document}